\documentclass[prb,groupedaddress,twocolumn,showkeys,showpacs,floatfix]{revtex4}
\usepackage{epsfig}
\usepackage{color}
\usepackage{amssymb,amsmath}
\begin{document}
\title{An electron motion induced by magnetic field pulse in bi-layer quantum wire}
\author{T. Chwiej}
\email{chwiej@fis.agh.edu.pl}
\affiliation{AGH University of Science and Technology, al. A. Mickiewicza 30, 30-059 Cracow,
Poland}
\begin{abstract}
\noindent
We consider theoretically the possibility of an electron acceleration in quantum wire by short
magnetic pulses lasted bewteen several to few tens of picoseconds.
We show that such possibility exists provided that, the electron is initially localized in part of
nanowire that consists of two vertically aligned layers which are tunnel coupled.
When a horizontally directed magnetic field, changeable in time, is also perpendicular to
the main axis of a wire, it generates a rotational electric field in it which pushes the upper and
the lower parts of the electron wavepacket in opposite directions. We have found however, that for
an asymmetric vertical confinement, the majority part of charge density starts to move in the
direction of local electric field in its layer but it also drags the minority part in the same
direction what results in coherent motion of an entire wavepacket. We discuss the dynamics of this
motion in dependence on the time characteristics of the magnetic pulse.
\end{abstract}
\keywords{quantum wire, ballistic transport, magnetic field}
\pacs{72.25.Dc,73.21.Hb}
\maketitle

\section{Introduction}

The tunnel coupling between two quantum wires has a great impact on the single electron transport
properties.\cite{eugster,bierwagen,lyo1,mourokh,fischer} If magnetic field penetrates such quantum
system  it can modify the magnitude of a tunnel coupling but the extent of such modifications
depends on the magnetic field strentgh and mutual arrangement of magnetic field and the wire axis.
\cite{lyo3,fischer4} If magnetic field is parallel to the wire axis, it squezees the wavefunctions
of magnetosubbands within each layer what results in lower value of a tunnel factor.
\cite{thomas,fischer4} 
On the other hand, if it is set perpendicularily to wire axis and to the layers coupled laterally
or vertically, it can hybridize the magnetosubbands between layers. This modifies, to a
large extent, the energy dispersion relation E(k) since the pseudo-gaps are opened and the
negative dispersion realation appear in energy spectrum.\cite{shi,lyo1} As we have shown in our last
paper, hybridization which leads to formation of pseudogaps can be utilized for tuning the magnitude
of spin polarization of wire's conductance provided that the wire has low density of
defects.\cite{chwiej2}

In present paper we study the dynamics of an electron motion in a bi-layer quantum wire made of
InGaAs/GaAs and GaAs/AlGaAs heterostructure which motion is induced by the magnetic pulse only.
Magnetic acceleration or deceleration of an electron in such nanostructure can be conducted provided
that: i) the wave functions originated from different layers are mutually hybridized, and, ii) the
time duration of magnetic pulse is between a few and tens of picoseconds.
At present, such short magnetic pulses can be generated by using the Auston's photoconductive
switches \cite{auston, spin_dynamics} or by the off-resonant magnetization of ferromagnetic thin
films with the terahertz laser pulses.\cite{offresonant}
Due to the Maxwell law i.e. $\partial\vec{B}/\partial t=-\nabla\times\vec{E}$, if the time
varying magnetic field is directed perpendicularily to a wire's axis as well as to
the vertically aligned transport layers, it generates the rotational electric field which try to
push two parts of the electron wavepacket being localized in both layers in the opposite directions.
The action of temporary rotational electric field on the electron wavepacket confined in vertical
bi-layer nanowire is schematically depicted in Fig.\ref{Fig:model}. We have investigated this
mechanism of acceleration of an electron in nanostructure of this kind and have found that the
assymetry introduced to the vertical confinement enables an entire electron wavepacket to move in a
particular direction but its further dynamics, i.e. after the pulse is finished, depends mainly on
the time length of magnetic pulse and on that, whether the magnetic field finally vanishes or its
value remains non-zero but is fixed for further times.
In the latter case, which is an analog of switching the magnetic field on, the
electron motion in constant magnetic field is governed by the magnetic force which for longer
times breaks the coherence between the upper and lower layers' parts of an electron wavepacket.

Paper is organized as follows. In Sec.\ref{Sec:model} we present numerical model of solving the
time-dependent Schr\"odinger equation for an electron confined in bi-layer nanowire, results of
simulations are presented and discussed in Sec.\ref{Sec:res}. We end up considerations with
conclusions given in Sec.\ref{Sec:con}.

\section{Theoretical model}
\label{Sec:model}

\begin{figure}[htbp!]
\hbox{
	\epsfxsize=70mm
       \rotatebox{0}{{ \epsfbox[0 652 473 842] {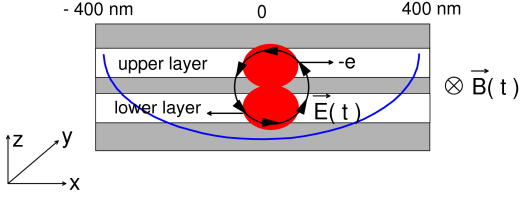}}}
        \hfill}
\caption{(Color online) The cross-section of a model bi-layer nanowire considered in paper.
The electron motion in x direction within harmonic oscillator potential (blue line) is stimulated
by rotational electric field to be induced by the time varying magnetic field which is pointed
along y direction, and, provided that both, 
upper and lower layers are tunnel coupled.}
\label{Fig:model}
\end{figure}

We start our considerations with the single electron Hamiltonian 
$\widehat{H}=(\widehat{\pmb{p}}+e\pmb{A})^2/2m^{*}+V_{c}(\pmb{r})$. Throuout the
paper we will use
the
time-dependent vector potential in non-symmetric form $\pmb{A}(t)=[z B(t),0,0]$ which gives
the magnetic field piercing the layers horizontally $\pmb{B}=[0,B(t),0]$ and being perpendicular to
the direction of an electron motion. We assume an electron can move along the
wire axis in x direction
within the harmonic oscillator potential ($V_{1}(x)=m^{*}\omega^2 x^2/2$) and
can tunnel between two
vertical layers which establish a double-well potential in z direction
($V_{2}(z)$). Its motion in y direction is frozen to the ground
state,  and is neglected in further discussion. Sketch of a model confinement potential is shown
in Fig.\ref{Fig:model}. Based on these assumptions, and additionally assuming
that the confinement potential in x and z directions is separable that is
$V_{c}(x,z)=V_{1}(x)+V_{2}(z)$, we make the problem simpler diagonalizing the part of Hamiltonian
dependent on z variable only $\widehat{h}_{z}=-(\hbar^{2}/2m^{*})\partial/\partial z^{2}+V_{2}(z)$.
From the set of its all eigenstates $f_{k}(z)$ we chose only two lowest eigenmodes to form the
functions basis $\{f_{1},f_{2}\}$. The rest lie much higher on energy scale and therefore their
contributions to final solution are neglected.
Next, calculating the matrix elements of Hamiltonian in this basis
$\widehat{H}_{k,k'}=\langle f_{k}|\widehat{H}|f_{k'}\rangle$ we get:
\begin{eqnarray}
 \nonumber
\widehat{H}_{k,k'}&=&\left(\widehat{T}_{x}+V_{1}(x)+E_{k}^{(z)}\right) \delta_{k,k'}\\
&+&\hbar\omega_{c}Z^{(1)}_{k,k'}\widehat{k}_{x}+\frac{m^{*}\omega^{2}_{c}}{2}Z^{(2)}_{k,k'}
\label{Eq:hkk}
\end{eqnarray}
where $\widehat{T}_{x}$ is kinetic operator for x direction, $E_{k}^{(z)}$ is a k-th eigenenergy
of Hamiltonian $\widehat{h}_{z}$, $\omega_{c}=eB/m^{*}$ is the cyclotron frequency,
matrix elements  $Z_{k,k'}^{(1)}$ and $Z_{k,k'}^{(2)}$ are defined as $Z_{k,k'}^{(m)}=\langle
f_{k}|z^{m}|f_{k'}\rangle$ and $\widehat{k}_{x}=\widehat{p}_{x}/\hbar=-i\partial/\partial x$.
Hamiltonian given in Eq.\ref{Eq:hkk} depends on time through the second and
the third terms in above equation which include $\omega_{c}$ because  magnetic
field change its value during an evolution of quantum system.
This form of energy operator allows us to define the wave function of an
electron as a two-component
object $| \Psi(x,t)\rangle=\sum_{k=1}^{2}\psi_{k}(x,t)|f_{k}\rangle$ for
which the
effective time-dependent
Schr\"odinger equation has the following form:
\begin{equation}
 i\hbar\frac{\partial}{\partial t}
 \left[
\begin{array}{c}
 \psi_{1}(x,t)\\
 \psi_{2}(x,t)\\
\end{array}
 \right]=
 \left[
\begin{array}{cc}
H_{11}&H_{12}\\
H_{21}&H_{22}
\end{array}
\right]
\left[
\begin{array}{c}
 \psi_{1}(x,t)\\
 \psi_{2}(x,t)\\
\end{array}
\right]
\label{Eq:sch}
\end{equation}
with matrix elements $H_{k,k'}$ defined by Eq.\ref{Eq:hkk}.

To describe the confinement in vertical direction within two layers we use the
following approximation
{\color{red}\cite{chwiej1,fischer}}:
\begin{equation}
V_{2}(z)=V_{max}\left\{ sin\left[(1+z/b)\pi/2 \right]+\alpha sin\left[ \pi(1+z/b) \right]\right\}
\label{Eq:vz}
\end{equation}
which is symmetric for $\alpha=0$ and the upper layers becomes deeper than the lower one for
$\alpha>0$. The latter case is studied in detail in next section.
The shape of $V_{2}(z)$ has a great impact on the dynamics of an electron in quantum wire.
For $\alpha=0$, basis wave functions $f_{1}(z)$  and $f_{2}(z)$ have defined parities. Then, the
third term in Eq.\ref{Eq:hkk} disappears and only the second term mixes both
components $\psi_{1}(x)$ and $\psi_{2}(x)$ of an electron's  wave function
through the off-diagonal elements $H_{12}$ and $H_{21}$ in Eq.\ref{Eq:sch} as the first term is
pure diagonal. On the other hand, when $\alpha>0$, the third term in Eq.\ref{Eq:hkk} also gives
contribution to off-diagonal elements in Eq.\ref{Eq:sch} and to the diagonal
ones as well.
Mixing of these components is crucial for inducing the motion of an electron by
the magnetic field pulse.
If there is no mixing during the time evolution, the
electron 
permanently occupy the lowest state that is $\psi_{1}(x,t)$ whereas
occupation of  $\psi_{2}(x,t)$ has to be zero
due to the condiction $E_{1}^{(z)}<E_{2}^{(z)}$, otherwise, mixing of these
elements should change the dynamics  of an electron
in the wire.
According to the definitions of the off-diagonal elements given by the second
and third terms in
Eq.\ref{Eq:hkk}, the mixing takes place only when $B\ne 0$ that is during the
magnetic pulse ($\partial B/\partial t \ne 0$) or for
constant magnetic field but provided that an electron moves within the wire.

Even though, the last two terms in Eq.\ref{Eq:hkk} can mix
the wave functions $\psi_{1}(x)$ and $\psi_{2}(x)$, only the second
term may force an electron to change its position. The proof is straightforward.
First, consider the ground state of electron for $B=0$, its wave function
at $t=0$ is given then by $\Psi(x,t)=[\psi_{1}(x,0),\psi_{2}(x,0)]=[g_{0}(x),0]$, with $g_{0}(x)$
being
the normalized square integrable function. If we now skip in Hamiltonian in Eq.\ref{Eq:sch} the term
$\hbar\omega_{c}Z_{k,k'}^{(1)}\widehat{k}_{x}$ and replace the time derivative with
the forward first-order finite difference formula, we obtain in first time step an approximated
expressions for: i) $\psi_{1}(x,\Delta t)=(1-i\Delta t
m^{*}\omega_{c}^{2}E_{1}/\hbar)\psi_{1}(x,0)$, where
$E_{1}$ is the total energy of electron at $t=0$ and the contribution from 
$\psi_{2}$ vanishes
since $\psi_{2}(x,0)=0$, and, ii) $\psi_{2}(x,\Delta t)=-(i\Delta t
m^{*}\omega_{c}^{2}Z_{21}^{(2)}/2)\psi_{1}(x,0)$. In
other words, part of $g_{0}(x)$  is moved from $\psi_{1}$ to $\psi_{2}$ and vice versa for further
time steps but the shape of $|\Psi(x,t)|^{2}$ is retained in any further time instant. Simply, this
Hamiltonian term is a diamagnetic energy shift which is responsible for optimizing the wave function
shape in z direction that is for its stronger localization in magnetic field. 
The dynamics of an electron during the time of simulation is governed
by the term $\hbar\omega_{c}Z_{k,k'}^{(1)}\widehat{k}_{x}$ which includes the derivative over x
variable.

We simulate the motion of an electron in bi-layer nanowire stimulated by the
magnetic field pulse by finding the solution of time-dependent Schr\"odinger equation
[Eq.\ref{Eq:sch}] in subsequent time instants by means of 4-th order Runge-Kutta method with time
step $\Delta t=10^{-4}\textrm{ ps}$.
For this purpose we have performed a set of numerical simulations on regular mesh of nodes  in x
direction. We limit the length of quantum wire to 800 nm and impose smooth confining potential in
this direction as
for the quantum oscillator $V_{1}(x)=m^{*}\omega_{0}^{2}x^2/2$ with oscillator strength
$\hbar\omega_{0}=0.5\textrm{ meV}$. This spatial constriction allows us to
study of an electron motion for quite long times after the magnetic pulse is
finished what is particularily important if an influence of magnetic force
($B_{y}=const$ ) on the coupling strength between the upper and lower parts of electron's
wavepacket is taken into account.
The basis functions $\{f_{1}(z),f_{2}(z)\}$ were also found numerically with
finite difference method after discretization of Hamiltonian $\widehat{h}_{z}$ on spatial mesh with
200 nodes for z direction. The number of nodes in x direction was set to 400.
All simulations were precedeed by diagonalization of Hamiltonian appearing in Eq.\ref{Eq:sch} to
prepare the starting wavepacket for $t=0$ i.e. to find $\Psi(x,t=0)=[\psi_{1}(x,0),\psi_{2}(x,0)]$.

Value of parameter b  which defines the extent of
vertical confinement in Eq.\ref{Eq:vz} was equal 30 nm. The height of tunnel barrier between the
upper and lower layers depends on magnitude of $V_{max}$. We set its
value equal to 150 meV for which the energy splitting between the bounding ($f_{1}$) and
the antibounding ($f_{2}$) basis states equals $E_{21}^{(z)}=E_{2}^{(z)}-E_{1}^{(z)}=2\textrm{
meV}$ for symmetric confinement ($\alpha=0$). When
$\alpha$ becomes non-zero then $\Delta E_{21}^{(z)}$ grows what eventually lowers the extent of
mixing $\psi_{1}$ and $\psi_{2}$ components of an electron's wave function. This can not be however
omitted since for $\alpha=0$, the
wavepackets localized in symmetric layers move in opposite directions [see Fig.\ref{Fig:model}] and
an electron wavepacket motion as a whole, in an arbitrarily chosen direction, can be obtain only for
$\alpha>0$ what will be shown below.
In paper we consider two cases, in first the electron is accelerated by single magnetic pulse, while
in second, its motion is induced when the magnetic field is switched on/off.

\section{Results}
\label{Sec:res}

\subsection{Acceleration of an electron by a single magnetic pulse}
\label{Sec:imp}
We start presentation of our results  for a case the electron is
accelerated by a single magnetic pulse defined as: $B(t)=B_{m}sin(\pi t/t_{imp})\cdot \Theta(t)\cdot
\Theta(t_{imp}-t)$, where $\Theta(t)$ is the Heaviside step function, the amplitude of magnetic
field equals $B_{m}=5\textrm{T}$ and the time duration of a pulse is changed within the range
$t_{imp}=5-50\textrm{ ps}$.
\begin{figure}[htbp!]
\hbox{
	\epsfxsize=80mm
       \rotatebox{0}{{ \epsfbox[0 305 569 842] {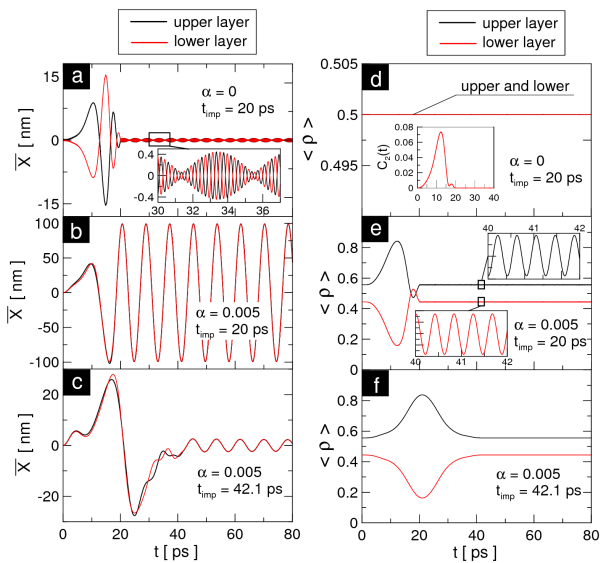}}}
        \hfill}
\caption{(Color online) Expectation value of electron position (left column) and the density (right
column) localized in the upper (red color) and in the lower layer (black color) during the time
of simulation for $\alpha=0$ (first row) and $\alpha=0.005$ (second and third rows).
Inset in figures (a) and (e) shows zooms of oscillations of $\bar{x}$ and $\langle
\rho\rangle$ remained after the magnetic pulse is ended.
The inset in (d) shows $C_{2}=\langle\psi_{2}|\psi_{2} \rangle $ that is an occupation of second
eigenstate for vertical quantization. Calculations were performed for parameters: $m^{*}=0.04$
(InGaAs) and $\Delta E_{21}^{z}=9.83\textrm{
meV}$ (a) and $\Delta E_{21}^{z}=9.9\textrm{meV}$ in (b) and (c).}
\label{Fig:rox}
\end{figure}
\begin{figure*}[htbp!]
\hbox{
	\epsfxsize=140mm
       \rotatebox{0}{{ \epsfbox[0 448 595 842] {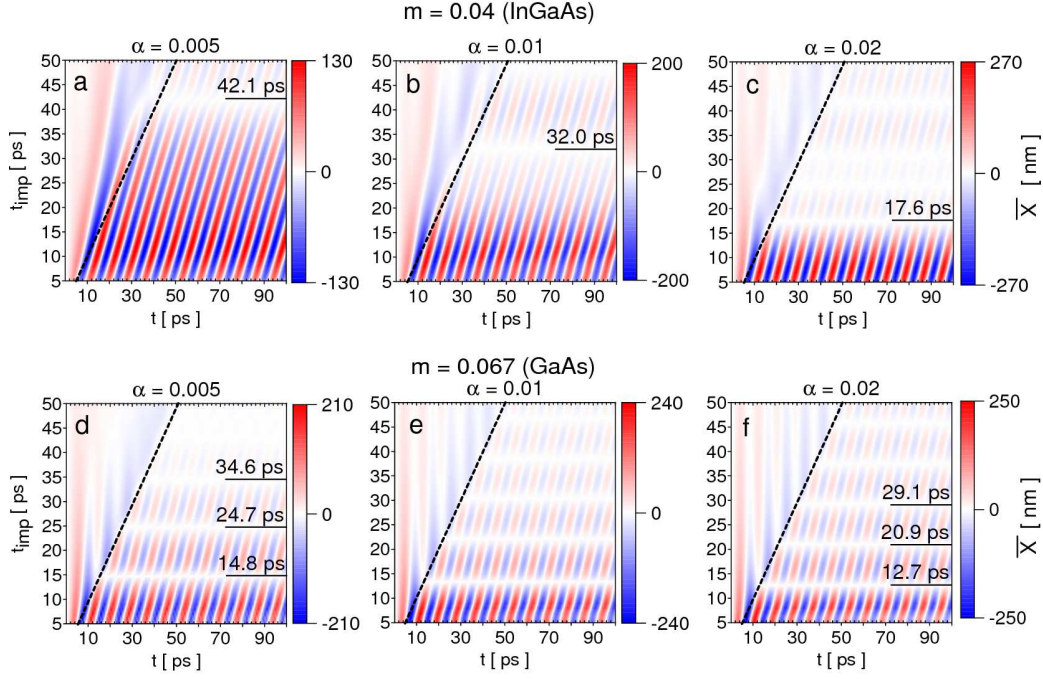}}}
        \hfill}
\caption{(Color online) Dependence of the normalized expectation value of electron position
($\bar{x}$) in bi-layer quantum wire  on time length of magnetic pulse and on the time of
simulation.
The inclined dashed lines mark the right border of magnetic field pulse. The upper row displays
the results obtained for InGaAs and effective mass $m_{InGaAs}^{*}=0.04$ while the lower one for
GaAs and effective mass $m_{GaAs}^{*}=0.067$. Parameter $\alpha$ which is shown on top of each
subfigure represents an actual asymmetry in vertical confinement.
}
\label{Fig:x}
\end{figure*}
Figure \ref{Fig:rox}(a) shows the time variations of the electron positions in
the upper and in the lower layers obtained according to Eqs. (\ref{Eq:xex}) and (\ref{Eq:xex2})
[see description below] for symmetric bi-layer nanowire ($\alpha=0$). The magnetic pulse
starts at $t=0$ and ends at $t=20\textrm{ ps}$.
Within this time interval, the parts of electron wavepacket localized in upper and in lower layers
oscillate in x direction  but with opposite phases i.e. the upper part starts to move to
the right while the second one, localized in the lower layer, to the left, and then, both
change the directions of their motions three times due to reflections within the regions of high
confining potential.
After the magnetic field eventually vanishes, these oscillations are strongly dumped but do not
disappear entirely. The inset in Fig.\ref{Fig:rox}(a) shows that even though for $t>t_{imp}$ the
amplitude of these oscillations become very small but oscillations remain stable for longer times.
The beating
pattern shown in this inset come into existence due to the overlap of two
kinds of oscilltions, one with period equal to $T_{1}=8.3\textrm{ ps}$ and
second with period $T_{2}=0.44\textrm{ ps}$. To determine their origin we first calculated the
expectation value of electron position in upper layer according to formula $\langle x
\rangle_{z>0}=\langle \Psi|x\cdot \Theta(z)|\Psi\rangle$. Exactly at the moment the magnetic pulse
ends ($B=0$), the electron wave function can
be expressed as $\Psi(x,y,z,t=t_{imp})=\sum_{k}\psi_{k}(x,t_{imp})f_{k}(z)g_{0}(y)$. The part of
wave
function which is dependent on x variable was expanded within the base constituted by Hermite
polynomials i.e. the quantum oscillator eigenstates $\Phi_{\mu}$:
\begin{eqnarray}
\nonumber
\Psi(x,y,z,t)&=&\sum_{k}\sum_{\mu}d_{k,\mu}\Phi_{\mu}(x)\\
&&\cdot f_{k}(z)g_{0}(y)e^{-i(E_{\mu}^{x}+E_{k}^{z}+E_{0}^{y})t/ \hbar}
\end{eqnarray}
In above equation, $d_{k,\mu}$ are the linear combination coefficients,
$\Phi_{\mu}(x)=C_{\mu}H_{\mu}(x)e^{-x^2}$ are the normalized Hermite polynomials
while three arguments appearing in phase factor are the eigenenergies for quantized motion of
electron in x ($E_{\mu}^{x}$), y ($E_{0}^{y}$) and z ($E_{k}^{z}$) directions.
According to this definition, the position of electron in an upper layer (dependence
on $g_{0}(y)$ and $E_{0}^{y}$ disappear after integration over y variable) reads:
\begin{eqnarray}
 \nonumber
\langle x \rangle_{z>0}=
\sum_{k,m}\sum_{\mu,\nu}&&
d_{k,\mu}^{*}d_{m,\nu}
\langle \Phi_{\mu}|x|\Phi_{\nu}\rangle
\langle f_{k}|\Theta(z)|f_{m}\rangle\\
&&\times e^{i\Delta E_{\mu,\nu}^{x}t/\hbar}
e^{i\Delta E_{k,m}^{z}t/\hbar}
\label{Eq:xex}
\end{eqnarray}
Since the amounts of probability density gathered in both layers may differ from unity we
normalized the expectation value of electron position in upper (or lower) layer:
\begin{equation}
\bar{x}_{z>0}=\frac{\langle x\rangle_{z>0}}{\langle\Psi|\Psi\rangle_{z>0}}
\label{Eq:xex2}
\end{equation}
The matrix elements $\langle \Phi_{\mu}|x|\Phi_{\nu}\rangle$ have non-zero values only if
$\mu=\nu\pm
1$ that is for $\Delta E_{\mu,\nu}=\pm \hbar\omega_{0}$ and the coupling of these non-diagonal terms
are responsible for the appearance of oscillations with longer period $T_{1}=h/\Delta E_{\mu,\mu\pm
1}=8.3\textrm{ ps}$, provided that,  $\Delta E_{km}^{z}$ vanishes for $k=m$ in Eq.\ref{Eq:xex}.
Otherwise, when conditions $k\ne m$ and $\mu=\nu\pm 1$ are fulfilled then the total phase factor
oscillates with period dependent on the sum of both energy differences i.e.
$\Delta E=\Delta E_{21}^{z}+\hbar\omega_{0}=10.33\textrm{ meV}$ what gives the period equal to
$T_{2}=0.44\textrm{ ps}$. However this value is very close to value $T_{2}^{'}=h/\Delta
E_{21}^{1}=0.42\textrm{ ps}$ what means that the coupling between vertical layers [due to the terms
$\langle f_{k}|\Theta(\pm z)|f_{m}\rangle$] is the main
source of high frequency oscillations visible in the inset in Fig.\ref{Fig:rox}(a).
Equation \ref{Eq:xex} can be further employed for calculation of $\langle\rho\rangle_{z>0}$ if 
the terms $\langle\Phi_{\mu}|x|\Phi_{\nu}\rangle$ will be substituted by
$\langle\Phi_{\mu}|\Phi_{\nu}\rangle=\delta_{\mu,\nu}$. This substitution eliminates dependence of
total phase on $\Delta E_{\mu,\nu}^{x}=0$ but leaves it still dependent on $\Delta E_{k,m}^{z}$
value. However in Fig.\ref{Fig:rox}(d) there are no oscillations at any time instant.
For $\alpha=0$, the amounts of electron denisty confined in both layers are equal, simply due
to the symmetry constriction imposed on the confining potential [see Fig.\ref{Fig:rox}(d)].

When $\alpha>0$ and the confinement energy in lower layer becomes higher than the one in an upper
layer, then for $t=0$ the larger part of electron density is gathered in upper layer [see
Figs.\ref{Fig:rox}(e) and (f)]. Therefore, when magnetic field starts to grow, the majority of
wavepacket move to right but this time the rest of wavepacket is also pulled into this direction. In
Figs. \ref{Fig:rox}(b) and \ref{Fig:rox}(c) we see that both densities move synchronously and 
independently on the time duration of magnetic pulse.  Due to an asymmetry that has been introduced
into the confining potential in vertical direction, the amounts of electron densities in both layers
are smoothly changed when $\partial B/\partial t\ne 0$ and for $t>t_{imp}$ become fixed [Figs.
\ref{Fig:rox}(e) and \ref{Fig:rox}(f)] besides the small oscillations for frequency
$\omega_{2}^{'}=2\pi/T_{2}^{'}$ visible in the inset in Fig.\ref{Fig:rox}(e).
The length of magnetic pulse has great impact on the amplitude of the electron oscillation within
quantum wire. For $t_{imp}=20\textrm{ ps}$ the amplitude of expectation value of electron position
in wire reaches even $\bar{x}_{max}=100\textrm{ nm}$ while for two times longer pulse the
amplitude of electron oscillations falls to value $\bar{x}_{max}=2.3\textrm{ nm}$. Despite the
different lengthes of these magnetic pulses, the period of oscillations  in both cases is the same
and equals $T_{1}=8.3\textrm{ ps}$ that is the dynamics of wavepacket is governed by the energy
levels structure of quantum harmonic oscillator established in x direction.
\newline
Analysis of the results showed in Fig.\ref{Fig:rox}(a-c) allows us to make a statement that the
characteristics of the electron motion induced by magnetic pulse in quantum bi-layer wire is
dependent on at least two factors: i) the time length of a magnetic pulse, and, ii) the degree of
asymmetry in vertical confinement provided that the amplitude of magnetic field is fixed.
In order to study the influence of these factors on dynamics of electron motion in the wire we
computed the expectation value of the electron position as function of $t_{imp}$ length and on the
time of simulation. Results for effective mass $m^{*}=0.04$ (InGaAs) are presented in first row in
Fig.\ref{Fig:x} for $\alpha=0.005,\ 0.01,\ 0.02$. These $\alpha's$ values give the energy
splittings between two lowest eigenstates in vertical direction equal to $\Delta
E_{21}^{z}=9.9\textrm{ meV},\ 10.1\textrm{ meV} \textrm{ and } 10.9\textrm{ meV}$.
For comparison, second row in this figure displays the results for $m^{*}=0.067$ (GaAs) and for
the energy splittings $\Delta E_{21}^{z}=2.3\textrm{ meV }, 3.1\textrm{ meV}\textrm{ and }
5.1\textrm{ meV}$. 
If an electron effective mass is small [first row in Fig.\ref{Fig:x}] then the increase of a
confining potential assymetry brings two main effects. First, the amplitude of $\bar{x}$ grows
[compare the ranges of  color scales in Figs.\ref{Fig:x}(a-c)] if value of $\alpha$ is increased.
For example, for $\alpha=0.005$ the amplitude of electron oscillations in x direction reaches
$\bar{x}_{max}=130\textrm{ ps}$ while for $\alpha=0.02$ it becomes two times larger.
Second, the time length of magnetic pulse which allows to obtain significant amplitudes
systematically decreases for larger $\alpha's$ values. For $\alpha=0.005$ the electron motion in
quantum wire can not be induced by pulses longer than $42.1\textrm{ ps}$ while for $\alpha=0.02$ it
is limited only to $17.6\textrm{ ps}$. For longer pulses, oscillations are still possible but
they might have much smaller amplitudes.

For larger effective mass [second row in Fig.\ref{Fig:x}] the upper limit for the time length of
magnetic pulse which still may induce large oscillations of electron position in bi-layer nanowire
is lowered to several picoseconds but its sensitivity to potential asymmetry is not so high as
in the previous case.
Now, its value decreases from $14.8\textrm{ps}$ to $12.7\textrm{ ps}$ when the
value of $\alpha$ increases from $0.005$ to $0.02$.
On the other hand, the largest amplitude of electron oscillations  becomes less dependent on
potential asymmetry for heavier particle, the change of $\alpha's$ value
from $0.005$ to $0.02$ increases the amplitude of $\bar{x}$ from 210 nm to 250 nm that is in
a three times thinner range than for the smaller effective mass.

\subsection{Electron motion induced by switching  the magnetic field on/off}
\label{Sec:imp2}

In this section we present the results obtained for a pulse generated by switching the magnetic
field on/off which is modelled in our simulations by following formula:
\begin{eqnarray}
\nonumber
B(t)&=&B_{m}sin(2\pi t/t_{imp}+\beta\pi/2)\cdot \Theta(t)\cdot \Theta(t_{imp}-t)\\
&+&\gamma\cdot B_{m}\cdot \Theta(t-t_{imp})
\label{Eq:imp2}
\end{eqnarray}
where: $\beta=0$ and $\gamma=1$ stand for the magnetic field growing in time [$\partial B/\partial
t>0$ for $t\le t_{imp}$] from 0 to $B_{m}$ while $\beta=1$ and $\gamma=0$ for vanishing field
[$\partial B/\partial t<0$ for $t\le t_{imp}$] from $B_{m}$ to 0.

\begin{figure}[htbp!]
\hbox{
	\epsfxsize=85mm
       \rotatebox{0}{{ \epsfbox[0 449 595 842] {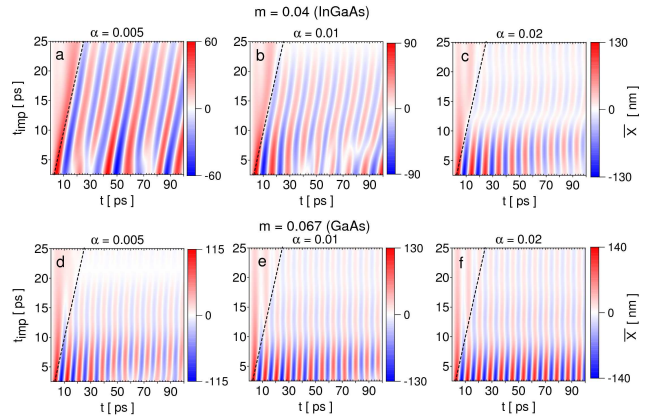}}}
        \hfill}
\caption{(Color online) Normalized expectation value of an electron position ($\bar{x}$) in an upper
layer depending on the time interval the magnetic field is switched on [$\partial B/\partial t>0$
for $t<t_{imp}$] and the time of simulation. Other parameters are given on the top of each
figure.}
\label{Fig:on}
\end{figure}

\begin{figure}[htbp!]
\hbox{
	\epsfxsize=80mm
       \rotatebox{0}{{ \epsfbox[0 523 203 842] {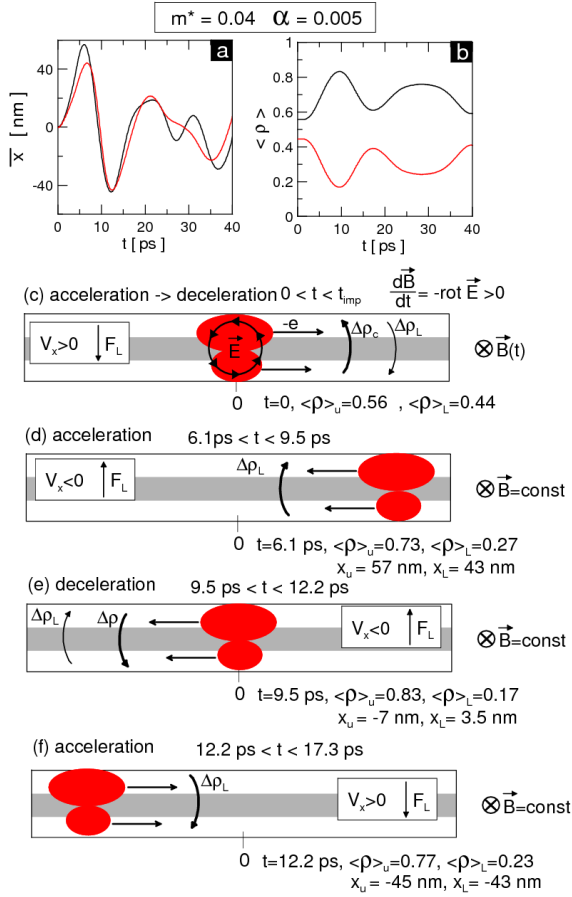}}}
        \hfill}
\caption{(Color online) The time variations of the expectation value of electron position (a) and
of the elctron density in the upper and in the lower layers (b) for $t_{imp}=5\textrm{
ps}$ and for magnetic field being switched on. Black and red colours in (a) and (b) mark the results
for the upper and lower layers, respectively. $\Delta \rho_{c}$ shows the density flow direction
caused by the stronger localization in the upper layer, $\Delta \rho_{L}$ indicate the density flow
due to magnetic force while $\Delta \rho$ marks the density flow caused by its unbalanced
accumulation in layers.}
\label{Fig:dud}
\end{figure}

\begin{figure}[htbp!]
\hbox{
	\epsfxsize=70mm
       \rotatebox{0}{{ \epsfbox[0 645 175 842] {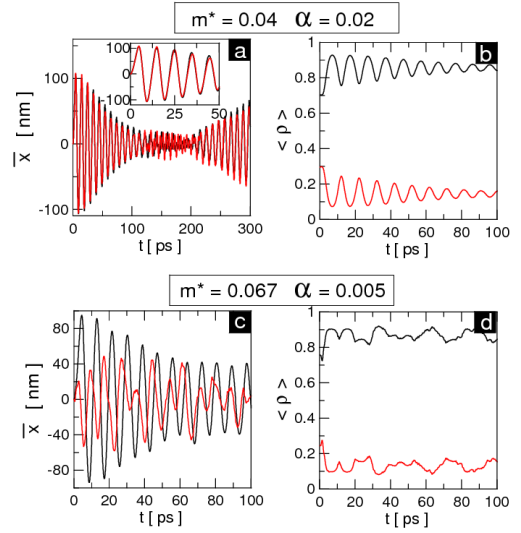}}}
        \hfill}
\caption{(Color online) The time changes of expectation values of electron positions (a,c) and of
the denisties (b,d) in the upper and in the lower layers for $t_{imp}=5\textrm{ ps}$ when magnetic
field is turned on. Black and red colors mark the results for an upper and for a lower  layer,
respectively. Values of $\alpha$ and of the effective masses are given on top of first and second
row.}
\label{Fig:dud4}
\end{figure}

Results for the first case, when magnetic field is turned on, are presented in  Fig.\ref{Fig:on}.
For InGaAs bi-layer wire, the dynamics of an electron changes qualitatively and quantitatively much
as the imbalance in the confining potential in the upper and in the lower layers raises [cf.
Figs. \ref{Fig:on}(a-c)]. For $\alpha=0.005$ and $\alpha=0.01$ the pattern of $\bar{x}$ is very
irregular for short pulses ($t<10\textrm{ps}$) and becomes regular that is with distinct single
period of oscillations for longer pulses. However, it does not concern the case with $\alpha=0.02$
[Fig.\ref{Fig:on}(c)] for which we may single out one main period being the same for all values of
$t_{imp}$. To explain the reasons of this irregularity we will analyze the changes in $\bar{x}$ and
in $\langle \rho \rangle$ in both layers for $t_{imp}=5\textrm{ ps}$, $\alpha=0.005$ and
$m^{*}=0.04$. Results are shown in Figs.\ref{Fig:dud}(a) and(b).

In Fig.\ref{Fig:dud}(a) we see that the time characterictics of $\bar{x}$ for both parts of 
wavepacket is similar for $t<24.2\textrm{ps}$ but thereafter their motions become decoupled even
though the denisty may still flow between the layers [Fig.\ref{Fig:dud}(b)]. At $t=0$ the electron
is at rest in the center of nanowire [Fig.\ref{Fig:dud}(c)] and when the magnetic field starts
raising the rotational electric field is generated  with opposite directions in upper and lower
layers. The upper layer confines a majority of electron
density which begins to move to the right according to the electric field. Simultaneously, it pulls 
the minority of density in this direction but against the x-component of
electric field in a lower layer. Even though the motion to the right implies the
magnetic force is directed downwards to the lower layer, in fact, density flows upwards since
the raising magnetic field enhances localization of density in deeper, the upper well, for
$t<5\textrm{ ps}$. Then the strength of magnetic field is fixed ($B=const$) and for time instant
$t=6.1\textrm{ ps}$ the wavepacket is stopped and after that turned back due to its reflection of
the high confinig potential on the right side [Fig.\ref{Fig:dud}(d)]. In the next stage the
electron moves to the left, first being accelerated ($6.1\textrm{ ps}<t<9.5\textrm{ ps}$) and next
when it passes the center of nanowire ($x=0$) being decelerated ($9.5\textrm{ ps}<t<12.2\textrm{
ps}$). Since the magnetic field is still applied to the system, the direction of magetic force is
now reversed [see Figs. \ref{Fig:dud}(d) and \ref{Fig:dud}(e)] what forces an additional part of
electron density to flow from lower to upper layer what we notice in Fig.\ref{Fig:dud}(b). At time
instant $t=12.2\textrm{ ps}$ the decelerated motion of electron is stopped on the left side of the
nanowire and next the directions of electron wavepacket motion and of magnetic force are again
reversed. For the right directed motion, the Lorenz force now easily shifts  the part of density
from the upper to lower layer. Thus, the existence of magnetic force is crucial for the dynamics of
the electron motion in bi-layer nanosystem because when the electron oscillates between the left and
right turning points, the density may flow bewteen layers but with vertical velocities what
eventually hinders the upper and lower parts of an electron wavepacket. For that reason 
the electron motion is constantly slowing down on average and both parts of
wavepacket, the upper and the lower ones, start to oscillate along x axis in asynchronous manner for
$t>24.2\textrm{ ps}$ [Fig.\ref{Fig:dud}(a)].

If the vertical confinement becomes more asymmetric ($\alpha=0.02$) or the effective mass is getting
larger ($m^{*}=0.067$) then an asynchronous motions of the upper and lower electron densities
also appear [see Figs. \ref{Fig:dud4}(a) and \ref{Fig:dud4}(c)]. In this case however, the
disproportion between the amounts of densities gathered in layers is significantly larger e.g.
$\langle\rho\rangle_{up}=0.7$ and $\langle\rho\rangle_{lo}=0.3$ for $t=0$ for $m^{*}=0.04$ and
$\alpha=0.02$ [Fig.\ref{Fig:dud4}(b)] and increases in time, what eventually has less impact on
phase perturbation in $\bar{x}$ oscillations in upper layer. The inset in Fig.\ref{Fig:dud4}(a)
shows that, the density flow  between layers for first few oscillations of $\bar{x}$ 
lowers only the amplitude of oscillations but motions of the upper and lower densities remain
synchronous. But when the amplitude gets smaller for $t>100\textrm{ ps}$, electron
slows down what in consequence lowers the magnitude of magnetic force and for certain time period
dumps the density flow between layers [compare the amplitudes of $\langle \rho\rangle$ at left and
right sides in Fig.\ref{Fig:dud4}(b)]. After that, the coupling between upper and lower 
denisties raises, both densities oscillate coherently what eventualy
makes the amplitude  of $\bar{x}$ to grow. This periodic coupling and decoupling of both densities
lead eventually to formation of beating pattern in $\bar{x}(t)$ function if an electron is allowed
to oscillate in nanowire for longer times [Fig.\ref{Fig:dud4}(a)].
Similar beating pattern for  $\bar{x}$ oscillations we also notice in
Fig.\ref{Fig:dud4}(c) for larger effective mass of an electron i.e. for $m^{*}=0.067$.
In this case, only a small assymetry in a confining potential ($\alpha=0.005$) is
required to localize about $70\%$ of total electron density in upper layer.
Unlike the previous case, now the motions of both parts of electron wavepacket are decoupled for any
time instant because as we see in Fig.\ref{Fig:dud4}(c), the oscillations of density in lower layer
exhibit, to some extent, a chaotic behaviour. Besides the fact, the density may still flow between
layers [see Fig.\ref{Fig:dud4}(d)], this chaotic motion of density in lower layer does not
perturb the frequency of $\bar{x}$ oscillations in upper layer but it influences their amplitude.

\begin{figure}[htbp!]
\hbox{
	\epsfxsize=80mm
       \rotatebox{0}{{ \epsfbox[0 115 571 842] {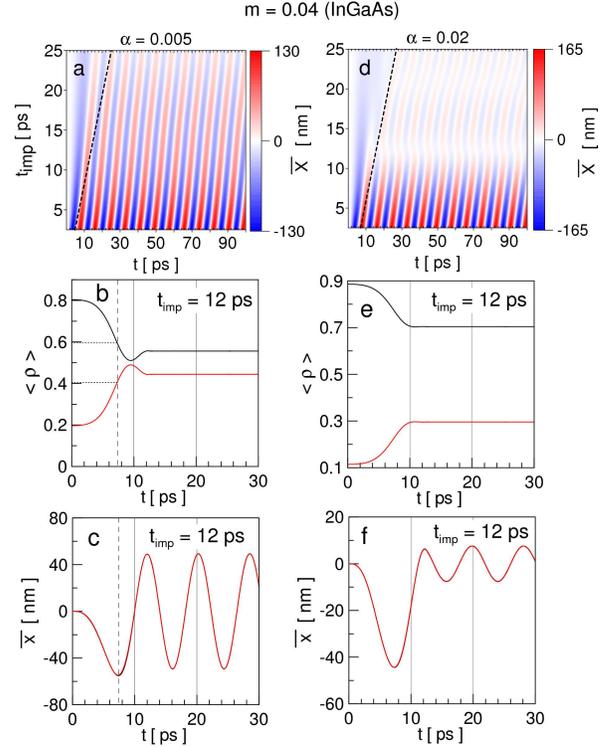}}}
        \hfill}
\caption{(Color online) (a,d) Dependence of $\bar{x}$ on the time length of magnetic pulse
and on the time of simulation for a case the magnetic field is switched off [$\partial B/\partial
t<0$ for $t<t_{imp}$ and $B=0$ for $t>t_{imp}$]. (b,e) The time changes of densities gathered in
upper layer (black color) and in lower layer (red color). (c,f) The normalized expectation value of
electron position in upper (black color) and in lower (red color) layers - they are the same. The
results showed in first and in second column were obtained for $\alpha=0.005$ and $\alpha=0.02$,
respectively.}
\label{Fig:off}
\end{figure}

In the last part of this section we  present the results of simulations performed for the pulse
being formed when the magnetic field is switch off [$\beta=1$ and $\gamma=0$ in Eq.\ref{Eq:imp2}].
Present case is different from that considered in Sec.\ref{Sec:imp} since now, the electron
wavepacket is prepared at $t=0$ for $B>0$ what means the densities confined in layers are
magnetically coupled at the beginning of simulation and their spatial localizations in vertical
direction are lowered when magnetic filed vanishes in time.
Results for $m^{*}=0.04$ are presented in Fig.\ref{Fig:off}.
Unlike the preceding case with magnetic field growing in time and being fixed afterwards, here we
have $B=0$ for $t>t_{imp}$ what implies the lack of an electron density flow between layers then.
Since the magnetic force is absent for $t>t_{imp}$, densities in both layers oscillate
coherently along the wire axis and the pattern of $\bar{x}$ does not change in time of
simulation i.e. the period of oscillations is fixed [cf. Figs. \ref{Fig:off}(a) and
\ref{Fig:off}(c)].
The degree of asymmetry in vertical confinement has large impact on sensitivity of the system on
a time length of magnetic pulse. For slightly asymmetric confinement ( $\alpha=0.005$) the amplitude
of $\bar{x}$ in upper layer is slowly diminishing as $t_{imp}$ grows. For larger asymmetry
($\alpha=0.02$), the amplitude of $\bar{x}$ gets smaller rapidly for $t_{imp}>12\textrm{ ps}$. This
qualitative difference in amplitude of oscillations of an electron position obviously
stems from the different dynamics of the electron density in layers for
$t<t_{imp}$ that is when the layers are magnetically  coupled through the
non-vanishing off-diagonal elements in Hamiltonian \ref{Eq:sch}. 
We will analyze this difference for $t_{imp}=12\textrm{ ps}$ for which the time variations of
$\langle \rho\rangle$ and $\bar{x}$ are shown in Figs. \ref{Fig:off}(b,e) and \ref{Fig:off}(c,f),
respectively.
In Fig.\ref{Fig:off}(b) we notice that for $\alpha=0.005$  about $80\%$ of electron density is
localized in upper layer at $t=0$. When the magnetic field is getting smaller, the electric
field generated in bi-layer wire pushes an upper, majority part of density to the left what drags
the lower one in this direction too. Both densities start to move coherently [see
Fig.\ref{Fig:off}(c)]. However, since the magnetic field becomes
weaker, the density is no longer strongly sqeezed in upper layer and it starts to flow to lower
layer what is additionally enhanced by the magnetic force which is directed downwards. At
$t=7.4\textrm{ ps}$, when the upper and lower densities are turned back, as large as $20\%$ of
total density has been transferred to the lower layer. After reversing the direction of 
velocity of an electron, the lower part of
wavepacket is now accelerated by electric field while the upper part is decelerated by it. For
this reason, the density still flows from an upper layer to  a lower one until it is almost equally
distributed [$\langle\rho\rangle_{u}=0.51$ and $\langle\rho\rangle_{l}=0.49$] for $t=9.5\textrm{
ps}$ [see Fig.\ref{Fig:off}(b)] what takes place just before the electron will pass the center of a
wire [cf. Figs. \ref{Fig:off}(b) and \ref{Fig:off}(c) for $t=0$].
In next two picoseconds, the direction of density flow is reversed, and when the magnetic field is
finally turned off about $56\%$ of total density is localized in upper layer and $44\%$ in lower
one. 

If distortion in vertical confinement is larger ($\alpha=0.02$), even $90\%$ of total density is
gathered in upper layer at $t=0$ and it drops to about $78\%$ immediately it reaches the left
turning point what one may notice in Fig.\ref{Fig:off}(e).
When the velocity of  wavepacket is reversed, the majority part of denisty is strongly decelerated
by an electric field what significantly diminishes the amplitude of oscillations of an electron
position along the wire's axis [Fig.\ref{Fig:off}(f)].
Therefore, we may conclude that for longer time intervals needed for switching the magnetic field
off, the dynamics of the majority part of electron density for $\alpha=0.02$ cannot adapt to the
time scale of the magnetic pulse what in consequence prevents the electron from gaining the large
values of $\bar{x}$ [see Fig.\ref{Fig:off}(d) for $t_{imp}>12\textrm{ ps}$].

\section{Conclusions}
\label{Sec:con}

The dynamics of an electron motion induced in quantum wire by a magnetic pulse was theoretically
studied by means of computer simulations. We have shown that such motion can be realized in a
semiconductor nanostructure which consist of two vertically stacked layers. If these layers are
tunnel coupled then the wave functions associated with two lowest bounding and antibounding
eigenstates for direction being perpendicular to both, the wire's axis and layers, can be easily
hybridized in magnetic field.
In such case, the rotational electric field generated due to the Maxwell law by
a non-zero time derivative of $\vec{B}$ may accelerate the charge density localized in the upper
and lower parts of bi-layer nanowire. For symmetric confinement in vertical direction, the upper and
lower charge densities are forced to move always in opposite directions according to the local
directions of the electric field. However, it was proven that the electron wavepacket, as a whole,
can be accelerated in arbitrarily chosen direction if the confinement in
upper and in lower layers becomes asymmetric. Then, the majority part of charge density localized in
one layer is pushed by the electric field and it simultaneously drags the minor part in the same
direction against the electric field generated in second layer. The dynamics of
an electron wavepacket becomes thus dependent on effective mass of an electron, degree of
asymmetry in the confining potential as well as on the
time characteristics of the magnetic pulse. It was found that generally only the
short time magnetic pulses with time duration of about several to maximally a few tens of
picoseconds may significantly change the motion energy of an electron. Moreover, the coherent
motion of both the upper and lower parts of wavepacket in the same direction can be
realized only if the magnetic coupling between layers vanishes when the magnetic pulse is ended.
This takes place when a single magnetic pulse is applied to the system or
a magnetic field is switched off. In the opposite case, the magnetic force
constantly changes the amounts of densities confined within layers what destroys their coherent
motion in a nanowire.

We think that discussed here the effect of temporary change of electron's motion energy by
picoseconds
magnetic pulses can successfully be applied to the nanostructures consisting of two
tunnel coupled wires allowing e.g. to sample the dynamical properties of many-body
interactions\cite{manybody} including the Wigner crystals.\cite{wigner}
Moreover, this effect can be also utilized in the nanostructures holding a two-dimensional electron
gas within the single, wide quantum well established in growth direction\cite{fischer4},
allowing for studies of the dynamical properties of electron transport in QPC\cite{qpc_tip,qpc2} or
temporary changes in the local potential landscape in conjunction with external biased
gates.\cite{qpc_tip}

\section*{Acknowledgements}
The work was financed by Polish Ministry of Science and Higher Education (MNiSW)
\section*{References}
\bibliography{lit3}
\end{document}